\documentclass[journal]{IEEEtran}

\usepackage{amsmath,amsfonts,amssymb}
\usepackage{graphicx}
\usepackage[caption=false,font=footnotesize]{subfig}
\usepackage{booktabs}
\usepackage{cite}
\usepackage{xurl}
\usepackage{stfloats}
\usepackage{mathtools}

\usepackage{xcolor}
\usepackage[
    colorlinks=true,
    citecolor=blue,
    linkcolor=blue,
    urlcolor=blue
]{hyperref}

\usepackage{multirow}

\renewcommand{\baselinestretch}{0.92}
\newcommand{\setR}{\mathcal{R}}
\newcommand{\setM}{\mathcal{M}}
\newcommand{\setN}{\mathcal{N}}
\newcommand{\setT}{\mathcal{T}}

\newcommand{\DT}{\Delta t}
\newcommand{\TTFTtag}{{\scriptscriptstyle\mathrm{TTFT}}}
\newcommand{\TPOTtag}{{\scriptscriptstyle\mathrm{TPOT}}}

\begin{document}

\title{Spatial LLM Workload Shifting Needs Foresight: Model Commitment for AI Data Center Operation under Power Grid Constraints}

\author{Bojun Du,~\IEEEmembership{Student Member,~IEEE},
        Hongyang Jia,~\IEEEmembership{Member,~IEEE},
        Tonghui Li,
        Qingchun Hou,~\IEEEmembership{Member,~IEEE},
        Ze Wang,
        Ershun Du,~\IEEEmembership{Member,~IEEE},
        and Ning Zhang,~\IEEEmembership{Senior Member,~IEEE}
        \vspace{-5ex}
\thanks{This work has been submitted to the IEEE for possible publication. Copyright may be transferred without notice, after which this version may no longer be accessible.}
\thanks{B. Du, H. Jia, E. Du, and N. Zhang are with the Department of Electrical Engineering, Tsinghua University, Beijing, China. Q. Hou is with the ZJU-UIUC Institute, Zhejiang University, Haining, China. T. Li and Z. Wang are with China Datang Technology Innovation Co., Ltd., Xiong'an, China.}%
}

\maketitle

\begin{abstract}
AI data centers may face power supply shortages during certain periods, requiring operators to shift large language model (LLM) inference workloads spatially to maintain service rates.
However, existing workload-shifting methods typically assume that any data center with sufficient computing resources can immediately serve shifted requests, which may lead to infeasible transfers and unserved demand.
This letter proposes model commitment (MC), a mixed-integer linear programming framework that jointly schedules model deployment and cross-site request routing under power constraints and electricity-price signals. 
First, MC formulates the intertemporal coupling introduced by model replica loading. Second, it translates prefill and decode latency requirements into the amount of demand that each replica can serve.
Case studies based on real-world data show that MC enables AI data center operators to achieve a 100\% service rate under time-varying grid conditions and reduce total operating cost by 29.0\%.
\end{abstract}


\section{Introduction}

\IEEEPARstart{D}{emand} for large language model (LLM) inference is growing rapidly, making access to sufficient electricity an emerging constraint on data-center operations \cite{IEA2025EnergyAI}. 
Therefore, AI data center operators need to coordinate workloads and computing resources under time-varying power supply constraints to improve service capability and operating efficiency.
To address these pressures, recent studies have used spatial workload shifting across geographically distributed data centers to relieve network congestion \cite{HierarchicalDCPowerNetworks2025}, and reduce operating costs through grid-aware coordination and local flexibility markets \cite{Dvorkin2025AgentCONCUR,SpatialDCFlexMarkets2026}. 
These approaches typically assume that any data center with sufficient computing resources can immediately serve shifted requests. However, an LLM inference request can be shifted only to where the corresponding model replica has been loaded. Ignoring this requirement may result in infeasible transfer decisions and leave requests unserved.

\begin{figure}[!t]
\centering
\includegraphics[width=0.85\columnwidth]{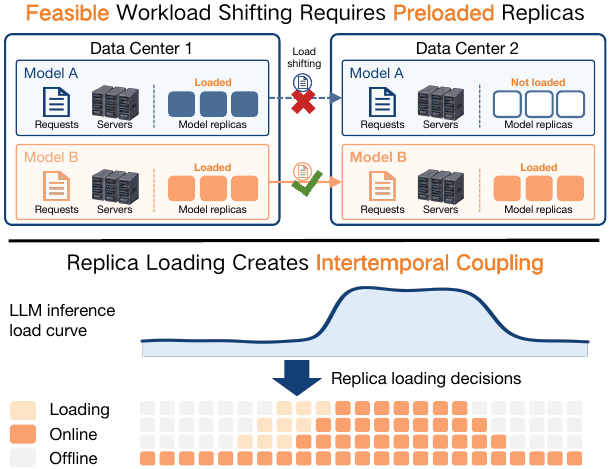}
\caption{Spatial LLM inference workload shifting requires service-ready replicas at the destination and advance loading before demand arrives.}
\label{fig:intro}
\end{figure}

As illustrated in Fig.~\ref{fig:intro}, feasible shifting of LLM inference needs the requested model replica to be loaded at the destination. 
Similar to generator start-up in unit commitment (UC), loading and warming up these replicas require substantial time. Thus, data center operators must decide model deployment in advance while coordinating computing resources under model-specific demand and power constraints \cite{ServerlessLLM2024}.
Once a replica is loaded and ready, latency requirements further determine how much demand it can serve. For LLM inference, these requirements include time to first token (TTFT) during prefill and time per output token (TPOT) during decode. However, existing service-level objective (SLO) formulations commonly constrain network transmission or task queueing delays \cite{CrossRegionalDCCapacity2026}.
Together, these gaps may lead to unserved requests and ignored latency constraints.

This letter proposes a model commitment (MC) framework for AI data center operation under power supply constraints and price signals. The main contributions are as follows. 
1) We formulate replica loading as an intertemporal commitment process and couple it with spatial request routing to improve service rates.
2) We establish the prefill and decode SLO constraints on the serving capacity of each committed replica to satisfy latency requirements.
Numerical results show that MC achieves a 100\% request-service rate and reduces operating cost by 29.0\%. Ablation studies further demonstrate that neglecting loading time or SLO limits leads to substantial unserved request demand.

\vspace{-1.5ex}

\section{Grid-Responsive LLM Inference Scheduling}
\label{sec:method}

MC is formulated from the perspective of AI data center operators to improve the service rate and reduce total cost. At a 15-min scheduling resolution, the operator takes request demand, power supply limits, and electricity prices as inputs. It determines the resident state of model replicas at each data center, and the spatial routing of requests across sites. 


\vspace{-1.5ex}

\subsection{Scheduling Objective}
\label{sec:obj}

Given model-specific token demand $D_{r,m,t}$, the requested token volume is either routed to a data center or left unserved:
\begin{equation}
\sum_n q_{r,m,n,t}+s_{r,m,t}=D_{r,m,t}, \ q_{r,m,n,t}\ge0,s_{r,m,t}\ge0,
\label{eq:routing}
\end{equation}
where $r\in\setR$ denotes demand regions, $m\in\setM$ denotes LLM types, $n\in\setN$ denotes data centers, and $t\in\setT$ denotes time periods. $q_{r,m,n,t}$ is the number of tokens routed to data center $n$, and $s_{r,m,t}$ is the unserved token volume.

Accordingly, MC minimizes the economic loss of unserved tokens and the total electricity cost:

\begin{align}
\min\quad
\sum_{r,m,t}\pi_m s_{r,m,t}+\sum_{n,t}\lambda_{n,t}P^{\mathrm{fac}}_{n,t}\DT,
\label{eq:objective}
\end{align}
where $\pi_m$ is the economic loss per unserved token for LLM type $m$, and $\lambda_{n,t}$ denotes nodal electricity price. $P^{\mathrm{fac}}_{n,t}$ is the facility power of data center $n$ at time $t$. 

\vspace{-5ex}

\subsection{Intertemporal Model Commitment and Resource Constraints}
\label{sec:mc}

Feasible routing requires the requested model to be loaded at the destination in advance, similar to generator start-up in UC. 
Let $y_{m,n,t}$ and $z_{m,n,t}$ denote the numbers of model-$m$ replicas whose loading is initiated and that are unloaded at data center $n$ in period $t$. Let $x_{m,n,t}$ denote the number of replicas that are fully service-ready throughout period $t$.
The replica state and its effective serving capacity are formulated as follows:
\begin{subequations}\label{eq:commitment}
\begin{flalign}
&x_{m,n,t}
=x_{m,n,t-1}
+y_{m,n,t-L^{\mathrm{ld}}_{m,n}}
-z_{m,n,t},
&&\label{eq:commitment_state}\\
&L^{\mathrm{ld}}_{m,n}
=\max\!\left\{
1,
\left\lceil
\frac{T^{\mathrm{rd}}_{m,n}}{\DT}
\right\rceil
\right\},\
\eta^{\mathrm{ld}}_{m,n}
=L^{\mathrm{ld}}_{m,n}
-\frac{T^{\mathrm{rd}}_{m,n}}{\DT},
&&\label{eq:loading_time}\\
&\widetilde{x}_{m,n,t}
=x_{m,n,t}
+\eta^{\mathrm{ld}}_{m,n}
y_{m,n,t-L^{\mathrm{ld}}_{m,n}+1},
&&\label{eq:effective_replicas} \\
&x_{m,n,t}, \ y_{m,n,t}, \ z_{m,n,t} \in \mathbb Z_{\ge0}
\end{flalign}
\end{subequations}
Eq.~\eqref{eq:commitment_state} updates the ready inventory by adding replicas whose loading was initiated $L^{\mathrm{ld}}_{m,n}$ periods earlier and subtracting unloaded
replicas. Eq.~\eqref{eq:loading_time} converts the physical readiness time $T^{\mathrm{rd}}_{m,n}$ into the integer loading horizon $L^{\mathrm{ld}}_{m,n}$. The residual
$\eta^{\mathrm{ld}}_{m,n}$ represents the fraction of the final loading period during which a newly ready replica can serve requests. Accordingly, Eq.~\eqref{eq:effective_replicas} defines the effective service-ready replica count $\widetilde{x}_{m,n,t}$ by combining the full-period ready inventory with this within-period contribution. For example, if $T^{\mathrm{rd}}_{m,n}=10$ min, it contributes $1/3$ to $\widetilde{x}_{m,n,t}$ in a 15-min period.

Loading and unloading cannot occur at the same time, and unavailable replicas cannot be unloaded. Hence,
\begin{equation}
\begin{alignedat}{2}
y_{m,n,t}
&\le \overline M \delta_{m,n,t},\;
z_{m,n,t}
\le \overline M (1-\delta_{m,n,t}),
\end{alignedat}
\label{eq:switching_feasibility}
\end{equation}
where $\delta_{m,n,t}\in\{0,1\}$ is the loading-mode indicator and $\overline M$ is a sufficiently large constant.

Replicas undergoing loading are excluded from $x_{m,n,t}$ but still occupy GPU resources. Thus, we define $\ell_{m,n,t}$ to count replicas that undergo loading during time $t$:
\begin{equation}
\ell_{m,n,t}=\sum_{k=0}^{L^{\mathrm{ld}}_{m,n}-1}y_{m,n,t-k},
\label{eq:loading_pipeline}
\end{equation}
Thus, GPU resources and memory constraints are modeled as 
\begin{subequations}\label{eq:computing_resources}
\begin{align}
&\sum_m a_{m,n}(x_{m,n,t}+\ell_{m,n,t})
\le A_{n,t}, \forall n,t &&\\
&\sum_m(g_m+\kappa_m)(x_{m,n,t}+\ell_{m,n,t})
\le G_{n,t}, \forall n,t&&
\end{align}
\end{subequations}
Where $a_{m,n}$ and $A_{n,t}$ are the GPUs per replica and available GPU count. The parameters $g_m$, $\kappa_m$, and $G_{n,t}$ are the model-weight memory, reserved KV-cache memory per replica, and available GPU memory, respectively.

MC further imposes the route-bandwidth constraint from \cite{CrossRegionalDCCapacity2026} on spatial request routing, and the data-center I/O constraint from \cite{Torpor2025} on model loading and unloading. The corresponding formulations are omitted for brevity.

The data-center power load is then derived from the computing resources used for model residency and request processing:
\begin{subequations}\label{eq:power}
\begin{align}
P^{\mathrm{idle}}_{n,t}&=\sum_m p^{\mathrm{idle}}_{m,n}(x_{m,n,t}+\ell_{m,n,t}),&&\\
P^{\mathrm{inf}}_{n,t}&=\frac{1}{\DT}\sum_{r,m}e_{m,n}q_{r,m,n,t},&&\\
P^{\mathrm{sw}}_{n,t}&=\frac{1}{\DT}\sum_m\left(E^{\mathrm{ld}}_{m,n}y_{m,n,t}+E^{\mathrm{ul}}_{m,n}z_{m,n,t}\right),&&\\
P^{\mathrm{fac}}_{n,t}&=\mathrm{PUE}_n
\big(P^{\mathrm{idle}}_{n,t}+P^{\mathrm{inf}}_{n,t}+P^{\mathrm{sw}}_{n,t}\big)
\le\overline P_{n,t}.&&
\end{align}
\end{subequations}
Here, $P^{\mathrm{idle}}_{n,t}$, $P^{\mathrm{inf}}_{n,t}$, and $P^{\mathrm{sw}}_{n,t}$ are idle, inference, and replica-loading powers, respectively.
The parameter $p^{\mathrm{idle}}_{m,n}$ is the base power per replica and $e_{m,n}$ is the incremental energy per output token. $E^{\mathrm{ld}}_{m,n}$ and $E^{\mathrm{ul}}_{m,n}$ denote loading and unloading energy. $\mathrm{PUE}_n$ denotes power usage effectiveness.
$\overline{P}_{n,t}$ specifies the power-consumption limit at data center $n$ during a demand-response event in period $t$.

\subsection{SLO-Constrained Serving Capacity Modeling}
\label{sec:latency}

A service-ready replica can accept routed demand only when both the prefill and decode latency SLOs are satisfied. Therefore, this subsection derives constraints on replica serving capacity from the TTFT and TPOT requirements.
During prefill, requests assigned to a replica share its service queue. Let $T^{\mathrm{pre}}_{m,n}$ denote the random prefill service time of model $m$ at data center $n$. Using an M/G/1 approximation \cite{She2026PLAServe}, the mean TTFT of one replica is
\begin{equation}
\mathbb{E}\!\left[\mathrm{TTFT}_{m,n}\right]
=
\overline{T^{\mathrm{pre}}_{m,n}}
+
\frac{
\nu_{m,n}\,
\overline{\left(T^{\mathrm{pre}}_{m,n}\right)^2}
}{
2\left(
1-\nu_{m,n}\overline{T^{\mathrm{pre}}_{m,n}}
\right)
},
\label{eq:mean_ttft}
\end{equation}
where $\nu_{m,n}$ is the request arrival rate assigned to one replica, $\overline{T^{\mathrm{pre}}_{m,n}}$ is the mean prefill service time.
The TTFT SLO requires the mean latency to remain below the limit $L_m^{\TTFTtag}$. Therefore, substituting Eq.~\eqref{eq:mean_ttft} into $\mathbb{E}[\mathrm{TTFT}_{m,n}] \le L_m^{\TTFTtag}$, SLO-constrained request rate satisfies:
\begin{equation}
\nu_{m,n}
\le
\Lambda^{\TTFTtag}_{m,n}
=
\gamma_m
\frac{
2\left(
L_m^{\TTFTtag}
-\overline{T^{\mathrm{pre}}_{m,n}}
\right)
}{
\overline{\left(T^{\mathrm{pre}}_{m,n}\right)^2}
+
2\left(
L_m^{\TTFTtag}
-\overline{T^{\mathrm{pre}}_{m,n}}
\right)
\overline{T^{\mathrm{pre}}_{m,n}}
},
\label{eq:ttft_capacity}
\end{equation}
where $\gamma_m\in(0,1]$ is an empirically calibrated safety factor that reserves capacity headroom for short-term fluctuations and approximation error of the mean queueing model \cite{MCParameterConfiguration}.

Given the average output length $\bar{o}_m$, the implied per-replica request arrival rate is
\begin{equation}
\nu_{m,n,t}
=
\frac{
\sum_r q_{r,m,n,t}
}{
\bar{o}_m\,
\widetilde{x}_{m,n,t}\,
\DT
}.
\label{eq:request_arrival_rate}
\end{equation}
Based on Eqs.~(\ref{eq:ttft_capacity})-(\ref{eq:request_arrival_rate}), the final TTFT constraint can be imposed in MC as
\begin{equation}
\sum_r q_{r,m,n,t}
\le
\Lambda^{\TTFTtag}_{m,n}
\bar{o}_m \cdot
\widetilde{x}_{m,n,t} \cdot
\DT.
\label{eq:ttft_routing_constraint}
\end{equation}

For the decode phase, each replica can sustain only part of the maximum throughput $R_m^{\max}$ to satisfy the TPOT latency limit $L_m^{\TPOTtag}$.
Profiling experiments show that the SLO-compliant fraction is approximately proportional to the TPOT limit within the practical operating range \cite{MCParameterConfiguration}. 
Therefore, the TPOT constraint is formulated as
\begin{equation}
\sum_r q_{r,m,n,t}
\le
\left(
\rho_m^{\TPOTtag}L_m^{\TPOTtag}
\right) \cdot
R_m^{\max} \cdot
\widetilde{x}_{m,n,t} \cdot \DT, 
\label{eq:tpot_capacity}
\end{equation}
where $\rho_m^{\TPOTtag}$ is an experimentally calibrated coefficient.

\section{Case Studies}

The case study considers three data centers and uses real-world nodal electricity-price profiles from three U.S. regions as grid signals. LLM request arrivals are based on real-world OpenAI workload traces \cite{wang2025burstgpt}. Requests for Qwen2.5-32B, Llama-3.3-70B, and DeepSeek-V3 are selected as representative workloads. Parameters such as PUE, model-serving throughput, and GPU specifications are calibrated from published measurements or experimental results. Complete parameter settings are provided in \cite{MCParameterConfiguration}.

\vspace{-2ex}

\begin{table}[!htbp]
\caption{Service and Cost Comparison.}
\label{tab:main}
\centering
\footnotesize
\setlength{\tabcolsep}{1pt}
\renewcommand{\arraystretch}{1.08}
\begin{tabular*}{\columnwidth}{
@{\extracolsep{\fill}}lcccc@{}
}
\toprule
\multirow{2}{*}{Method}
& \multicolumn{1}{c}{Service}
& \multicolumn{1}{c}{\shortstack{Unserved-token}}
& \multicolumn{1}{c}{Total cost}
& \multicolumn{1}{c}{Elec. cost} \\
& \multicolumn{1}{c}{rate (\%)}
& \multicolumn{1}{c}{penalty (k\$)}
& \multicolumn{1}{c}{(k\$)}
& \multicolumn{1}{c}{(\$/10$^9$ tokens)} \\
\midrule
B1-Myopic        &  95.43 &  21.78 &  31.32 & 33.05 \\
B2-Static        &  97.66 &   5.93 &  21.05 & 51.21 \\
B3-ZeroLeadTime  &  67.40 & 129.93 & 134.85 & 24.10 \\
B4-NoPriceSignal & 100.00 &   0.00 &  12.56 & 41.53 \\
B5-NoTTFT        &  56.89 & 204.08 & 206.97 & 16.77 \\
B6-NoTPOT        &  78.04 &  53.16 &  61.15 & 33.86 \\
\textbf{Proposed MC}
& 100.00
& 0.00
& 8.92
& 29.48 \\
\bottomrule
\vspace{0.2pt}
\end{tabular*}
\parbox{\columnwidth}{\footnotesize\raggedright
\emph{Note:} B1 uses a 30-min look-ahead without full-horizon information; B2 fixes replica placement; 
B3 neglects loading time; B4 ignores nodal price differences; 
B5 and B6 omit TTFT and TPOT constraints, respectively. 
B3, B5, and B6 are physically replayed under the full constraints.
}
\end{table}


Table~\ref{tab:main} compares the service rate and cost of MC with two operational baselines and four ablation variants. 
Compared with B1-Myopic and B2-Static, MC reduces total cost and increases the service rate to 100\%. Among the ablations, B4-NoPriceSignal retains full service but increases cost, whereas B3-ZeroLeadTime, B5-NoTTFT, and B6-NoTPOT have substantially lower service rates. 
MC has a higher average electricity cost than B3 and B5, primarily because it serves much more demand. Achieving full service requires additional capacity at higher-priced data centers, thereby increasing the marginal electricity cost.
These results indicate that feasible LLM inference workload shifting must account for model-residency dynamics. Such shifting also needs to satisfy serving constraints on throughput and latency. MC meets these requirements under power supply constraints, effectively improving the service rate.

\begin{figure}[!htbp]
\centering
\includegraphics[width=0.9\columnwidth]{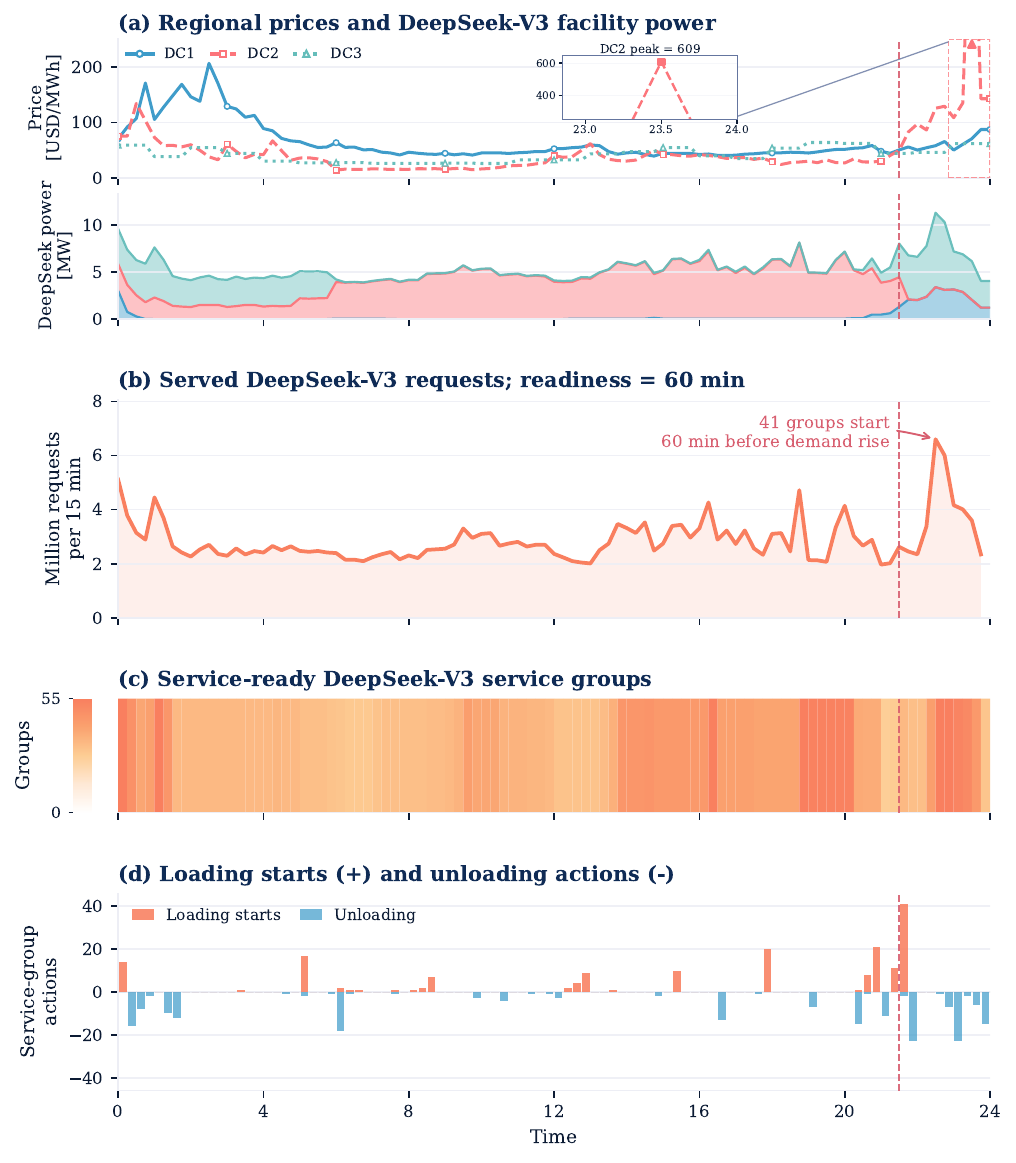}
\caption{DeepSeek-V3 power consumption and commitment actions under MC.}
\label{fig:commitment}
\end{figure}

Fig.~\ref{fig:commitment} presents the DeepSeek-V3 serving and commitment decisions obtained by MC. The comparison of panels (a) and (b) shows that facility power varies less sharply than the request curve. Unloading replicas reduces immediate power consumption, but restoring the released capacity requires lead time and adds switching costs. Therefore, MC retains some resident capacity when demand declines rather than repeatedly cycling replicas, resulting in a persistent base load.

Panels (c) and (d) further illustrate the commitment actions underlying the observed power profile. Although request demand peaks near 23:00, MC begins reconfiguring the deployment around 21:30, loading additional replicas at DC1 and DC3 before unloading those at DC2. The supply shortage and rising electricity price at DC2 drive this anticipatory reallocation toward DC1 and DC3.
These results show that MC prepares service-ready capacity before the demand peak and reallocates it in response to power supply constraints and regional price changes.

\section{Conclusion}

This letter proposes MC for workload shifting of AI data centers under time-varying grid conditions. MC coordinates intertemporal model commitment and request routing under LLM-specific latency limits. By explicitly modeling these constraints, MC ensures feasible LLM inference workload shifting and achieves a 100\% service rate with a 29.0\% reduction in total operating cost.

\vspace{-1ex}

\bibliographystyle{IEEEtran}
\bibliography{references}

@techreport{IEA2025EnergyAI,
  author      = {{International Energy Agency}},
  title       = {Energy and {AI}},
  institution = {IEA},
  address     = {Paris},
  year        = {2025},
  url         = {https://www.iea.org/reports/energy-and-ai}
}

@article{HierarchicalDCPowerNetworks2025,
  author  = {Liu, Junhong and Teng, Fei and Hou, Francis Yunhe},
  title   = {Synergising Hierarchical Data Centers and Power Networks: A Privacy-Preserving Approach},
  journal = {IEEE Transactions on Smart Grid},
  year    = {2025},
  month   = {11},
  volume  = {16},
  number  = {6},
  pages   = {5083--5098},

}

@inproceedings{ServerlessLLM2024,
  author    = {Fu, Yao and Xue, Leyang and Huang, Yeqi and Brabete, Andrei-Octavian and Ustiugov, Dmitrii and Patel, Yuvraj and Mai, Luo},
  title     = {{ServerlessLLM}: Low-Latency Serverless Inference for Large Language Models},
  booktitle = {18th USENIX Symposium on Operating Systems Design and Implementation (OSDI 24)},
  year      = {2024},
  month     = {July},
  address   = {Santa Clara, CA},
  pages     = {135--153},
  publisher = {USENIX Association},
}

@inproceedings{wang2025burstgpt,
  title={Burstgpt: A real-world workload dataset to optimize llm serving systems},
  author={Wang, Yuxin and Chen, Yuhan and Li, Zeyu and Kang, Xueze and Fang, Yuchu and Zhou, Yeju and Zheng, Yang and Tang, Zhenheng and He, Xin and Guo, Rui and others},
  booktitle={Proceedings of the 31st ACM SIGKDD Conference on Knowledge Discovery and Data Mining V. 2},
  pages={5831--5841},
  year={2025}
}

@article{Dvorkin2025AgentCONCUR,
  author  = {Dvorkin, Vladimir},
  title   = {Agent Coordination via Contextual Regression ({AgentCONCUR}) for Data Center Flexibility},
  journal = {IEEE Transactions on Power Systems},
  year    = {2025},
  volume  = {40},
  number  = {2},
  pages   = {1832--1842},
  month   = {March},
}

@article{SpatialDCFlexMarkets2026,
  author  = {Chen, Boyu and Che, Yanbo and Takc{\i}, Mehmet T{\"u}rker
             and Qadrdan, Meysam and Zhou, Yue},
  title   = {Spatial Flexibility Provision From Geographically Dispersed
             Data Centers Enabling Coordinated Operation of Multi-Local
             Flexibility Markets},
  journal = {IEEE Transactions on Smart Grid},
  year    = {2026},
  volume  = {17},
  number  = {2},
  pages   = {1371--1381},
}

@article{CrossRegionalDCCapacity2026,
  author  = {Han, Jianpei and Du, Ershun and Du, Bojun and Li, Yaowang and Guo, Jingrong and Zhang, Ning and Kang, Chongqing},
  title   = {Evaluating the Dispatchable Capacity of Cross-Regional Data Center Clusters Toward Power System Operation},
  journal = {IEEE Transactions on Smart Grid},
  year    = {2026},
  month   = mar,
  volume  = {17},
  number  = {2},
  pages   = {1180--1193},
  doi     = {10.1109/TSG.2025.3623107}
}

@inproceedings{Torpor2025,
  author    = {Minchen Yu and Ao Wang and Dong Chen and Haoxuan Yu
               and Xiaonan Luo and Zhuohao Li and Wei Wang
               and Ruichuan Chen and Dapeng Nie and Haoran Yang
               and Yu Ding},
  title     = {Torpor: {GPU-Enabled} Serverless Computing for
               {Low-Latency}, {Resource-Efficient} Inference},
  booktitle = {2025 USENIX Annual Technical Conference
               (USENIX ATC 25)},
  year      = {2025},
  pages     = {597--612},
  publisher = {USENIX Association},
  url       = {https://www.usenix.org/conference/atc25/presentation/yu}
}

@inproceedings{She2026PLAServe,
  author    = {Jianshu She and Zonghang Li and Hongchao Du and Shangyu Wu
               and Wenhao Zheng and Eric P. Xing and Zhengzhong Liu
               and Huaxiu Yao and Jason Xue and Qirong Ho},
  title     = {{PLA-Serve}: A Prefill-Length-Aware {LLM} Serving System},
  booktitle = {Proceedings of Machine Learning and Systems},
  volume    = {8},
  pages     = {2096--2112},
  year      = {2026},
  publisher = {MLSys},
}

@misc{MCParameterConfiguration,
  title        = {Detailed Parameter Settings for the Case Study},
  url          = {https://github.com/EilabDBJ/Model-Commitment/blob/main/parameter_configuration.md}
}

\end{document}